\documentclass[aps,prl,floats,floatfix,showpacs,twoside,preprintnumbers,superscriptaddress,nofootinbib]{revtex4}

\usepackage{graphicx}

\begin{document}

\title{Preferential Attachment in the Interaction between\\ 
Dynamically Generated Interdependent Networks}

\author{Boris~Podobnik}
\affiliation{Center for Polymer Studies and Department of Physics, Boston 
University, Boston, MA 02215, USA}
\affiliation{Faculty of Civil Engineering, University of Rijeka, 51000 
Rijeka, Croatia}
\affiliation{Faculty of Economics, University of Ljubljana,
  1000 Ljubljana, Slovenia}

\author{Davor Horvati\'c}
\affiliation{Physics Department, Faculty of Science, University of Zagreb, 
10000 Zagreb, Croatia}

\author{Mark Dickison}
\affiliation{Center for Polymer Studies and Department of Physics, Boston 
University, Boston, MA 02215, USA}

\author{H.~Eugene~Stanley}
\affiliation{Center for Polymer Studies and Department of Physics, Boston 
University, Boston, MA 02215, USA}

\begin{abstract}
We generalize the scale-free network model of Barab\`asi and Albert
[Science 286, 509 (1999)] by proposing a class of stochastic models for
scale-free interdependent networks in which interdependent nodes are not
randomly connected but rather are connected via preferential attachment
(PA). Each network grows through the continuous addition of new nodes,
and new nodes in each network attach preferentially and simultaneously
to (a) well-connected nodes within the same network and (b)
well-connected nodes in other networks.  We present analytic solutions
for the power-law exponents as functions of the number of links both
between networks and within networks. We show that a cross-clustering
coefficient vs. size of network $N$ follows a power law.  We illustrate
the models using selected examples from the Internet and finance.
\end{abstract}


\maketitle

Network research is a topic of interest with many applications in
physics. For example, in quantum chromodynamics, network models have
been used in calculating quark-hadron transition parameters
\cite{Reeves}, and Bose-Einstein condensation has connections with
network theory \cite{Bianconi}. Scale-free behavior has been observed in
a huge variety of different networks, ranging from the Internet to
biological networks
\cite{Faloutsos,Barabasi99,Adamic,Albert00,Milo,Garlaschelli03,AlbertRMP,Colizza,Onnela,Liben,Borgatti,Plamen12}.
With few exceptions
\cite{Buldyrev10,VespignaniN,ParshaniPNAS,Gao2011,Buld11,Shao11,Son11,GaoNP11},
most network studies have focused on single networks that neither
interact with nor depend on other networks \cite{AlbertRMP}. Recently it
was noted \cite{ParshaniEPL} that port and airport networks interact
with each other and that the coupling between these networks is not
random but correlated. Our general assumption is that real-life
scale-free networks are correlated rather than isolated, and that
preferential attachment (PA) and its variants
\cite{Krapivsky00,Dorogovtsev00,AlbertPRL00} control not only the
dynamics within a network but also the dynamics between different
networks. In bank-insurance firm networks, for example, we expect large
banks to be more attractive to insurance firms than small banks.

Recently, Ref.~\cite{Holyst06} investigated the behavior of the Ising
model on two connected Barab\`asi -- Albert networks in which each node of
the network has a spin, and $J_{AB} =J_{BA}$ are the coupling constants
between spins in different networks.  Here we emphasize scale-free
interdependent network dynamics rather than the robustness of
interacting networks
\cite{Buldyrev10,ParshaniPNAS,Gao2011,Buld11,Shao11,Son11,GaoNP11}.  We
propose a class of stochastic models for scale-free interdependent
networks in which interdependent nodes are not randomly connected but
are the result of PA.  In our approach PA controls not only the dynamics
of each network but also the interaction between different networks.
First, we define a coupled Barab\`asi-Albert (BA) model {\bf I} composed
of two interdependent networks BA$_1$ and BA$_2$ where the PA between
different networks and within a network is identical.  Second, we define
a coupled BA model {\bf II} where the PAs between different networks and
within a network are distinct. Third, we define a ``network of networks"
model.  Finally, we present two examples of interdependent networks,
from the Internet and from finance.


There are many interdependent networks or ``networks of networks" (NON)
in real-world data \cite{Gao2011}. For example, in physiology, the human
body is an example of a NON system that includes the respiratory,
nervous, and cardiovascular systems.
 
As an example of a NON we consider the Internet: a network of routers or
autonomous systems (AS) connected by links
\cite{BarabasiPNAS02,Shavitt}. Using the fractal concept in which each
part of a complex system is an approximate reduced-size copy of the
whole---i.e., is ``self-similar''---we analyze AS connections not for
the entire world \cite{BarabasiPNAS02,Shavitt} but rather the Internet
connections between three countries.  Specifically, we study AS
connections between the US, Germany, and the UK recorded over an
18-month period.  For each of the three countries we study both total
connectivity ($k^T$) and the number of links ($k$) within each country.

In the following analyses, in order to estimate power-law exponent
$\gamma$ for a power-law distributed variable $k$ with $P(k)\propto
k^{-\gamma}$ we apply two methods. In the Zipf ranking approach
in which $R$ denotes rank, one commonly applies the regression
\begin{equation}
\log(\mathrm{k})=a- \zeta ~\log(R),
\label{zipf} 
\end{equation}
where $\zeta = 1/(\gamma-1)$, 
which is strongly biased in small samples
\cite{Newman05,Newman09,Gabaix11}. In the first method, in order to
overcome this bias, we apply a recently-proposed regression method
\cite{Gabaix11}
\begin{equation}
 \log(\mathrm{R}-1/2)=a- (\hat \gamma - 1) ~  \log(k).
\label{gabaix} 
\end{equation}
In the second method we estimate the power-law exponent $\hat
\gamma'$ using the equation
\begin{equation}
 \hat \gamma' = 1 + N [ \Sigma_{t=1}^{N}
  \log (k_t/k_{\rm min}) ]^{-1}, 
  \label{alpha}
\end{equation}
where $k_{\rm min}$ is the smallest value of $k_t$ for which the
power-law behavior holds, and the sum runs only over those values of
$k_t$ that exceed $k_{\rm min}$
\cite{Newman05,Newman09}. Equation~(\ref{alpha}) is equivalent to the
well-known Hill estimator where the standard error on $\hat \gamma$,
which is derived from the width of the likelihood maximum, is $\sigma =
\frac{\gamma' - 1}{\sqrt{n}} + O(1/n)$.

\begin{figure}
\centering
\includegraphics[width=0.5\columnwidth]{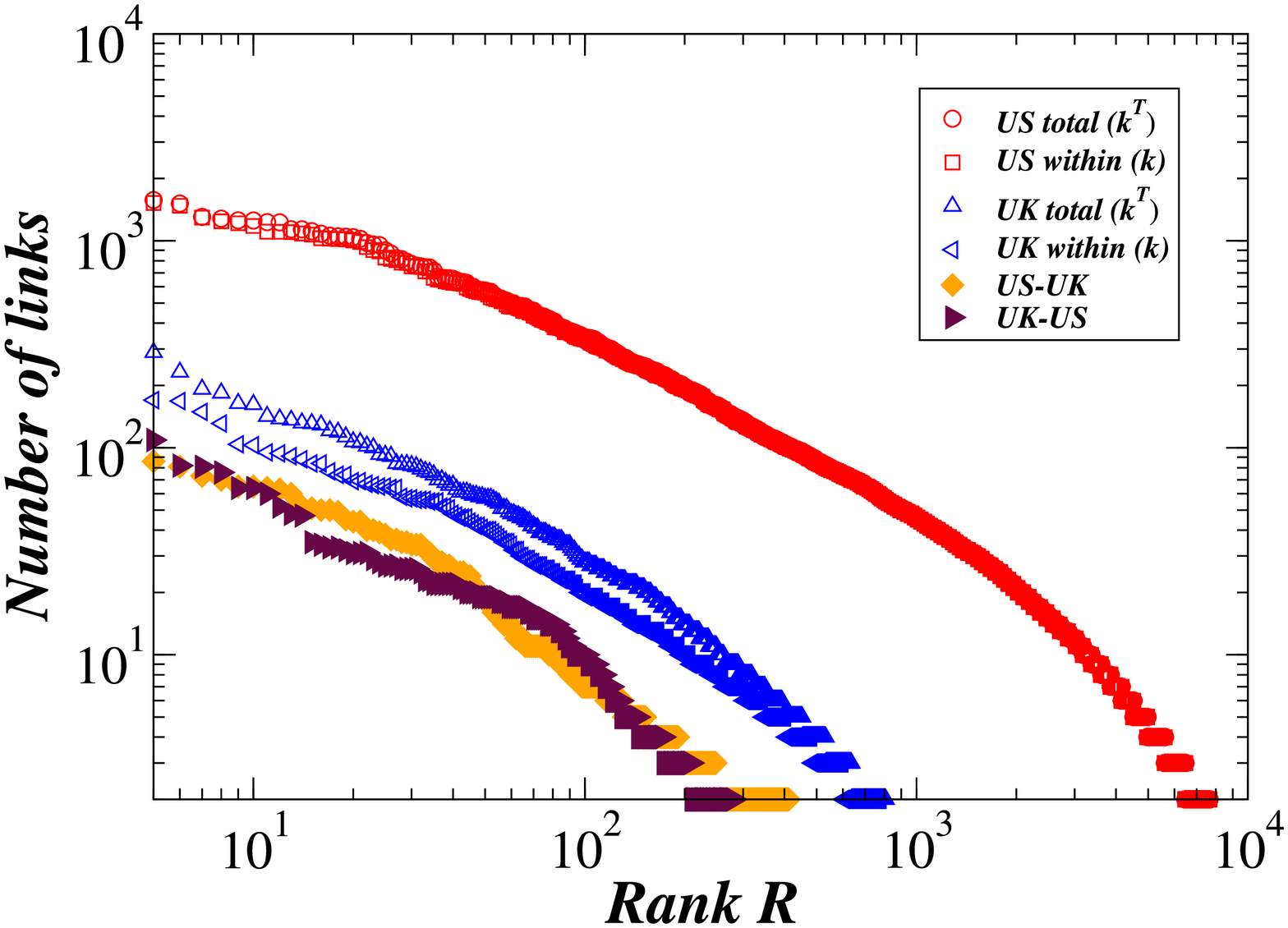}
\caption{Level of interdependency between countries (no interdependency,
 no  cross-links).  
 Approximate power laws in the Internet obtained for the  plot
 of  number of links versus rank $R$ 
 in AS interdependent networks between different countries. 
 We calculate the exponents of Eqs.~(\ref{gabaix}) and (\ref{alpha})  for
  the total number of links and the number of links
  established only within  each country. For the US we obtain  
$\hat \gamma^T = 2.24 \pm 0.01  ~(\hat \gamma'^T = 2.17 \pm 0.04 )  $ and 
$\hat \gamma = 2.26 \pm 0.01  ~(\hat \gamma' = 2.17 \pm 0.04)$,  
and for the UK   
$\hat \gamma^T = 2.00 \pm 0.01  ~(\hat \gamma'^T = 2.21 \pm 0.11 )  $ and 
$\hat \gamma = 2.06 \pm 0.01  ~(\hat \gamma' = 2.20 \pm 0.11)$. 
 For the cross-links, we obtain  for US-UK $\hat \gamma = 2.04 
 \pm 0.03  ~(\hat \gamma' = 1.98 \pm 0.09  )$
  and for UK-US  $\hat \gamma = 2.39 \pm 0.02  
  ~(\hat \gamma' = 2.54 \pm 0.24 )$.  
  }
\label{fig1}
\end{figure}

For the sake of simplicity, Fig.~\ref{fig1} shows the NON results of our
study on network of routers for only two interdependent countries, the
US and the UK. We find that 9685 cities in the US and 1170 cities in the
UK are connected by routers. For each country we show (a) the number of
links established within the country, (b) the total number of links
established not only within the country but also with the coupled
country, and (c) the cross links, e.g., the links established from the
UK routers to the US routers, and vice versa. Note that no cross-links
between UK and the US router networks implies no interdependency between
the networks. We find that each Zipf plot of $k$ in Eq.~\ref{zipf}
exhibits an approximate power-law scaling. For each country we find that
$\hat \gamma^T$ obtained for total connectivity is smaller than
$\hat\gamma$ obtained for links within a single country---employing
Eqs.~(\ref{gabaix})--(\ref{alpha}) for the US we find $\hat \gamma^T =
2.24 \pm 0.01 ~(\hat \gamma'^T = 2.17 \pm 0.04 ) $ and $\hat \gamma =
2.26 \pm 0.01 ~(\hat \gamma' = 2.17 \pm 0.04)$.  For the UK we find
$\hat \gamma^T = 2.0 \pm 0.01 ~(\hat \gamma'^T = 2.21 \pm 0.11 ) $ and
$\hat \gamma = 2.06 \pm 0.01 ~(\hat \gamma' = 2.20 \pm 0.11)$.  We note
that similar results for the exponents of degree distributions do not
imply that interdependency exists between two networks. To this end, for
the cross-links which quantify the level of interdependency between
countries (again, no interdependency, no cross-links), we find for US-UK
$\hat \gamma = 2.04 \pm 0.03 ~(\hat \gamma' = 1.98 \pm 0.09 )$ and for
UK-US $\hat \gamma = 2.39 \pm 0.02 ~(\hat \gamma' = 2.54 \pm 0.24 )$.
We also show the cross-link interdependent router connections between
the UK and Germany, with 1170 cities in the UK and 1989 cities in the
Germany.  We find for Germany-UK $\hat \gamma = 2.01 \pm 0.03 ~(\hat
\gamma' = 2.01 \pm 0.15 )$ and for UK-Germany $\hat \gamma = 2.51 \pm
0.05 ~(\hat \gamma' = 2.20 \pm 0.25)$. Note that the similar degree
distributions shown in Fig.~1 never guarantee similar mechanisms of
network generations or even other characteristics of networks such as
community structures and degree assortativity \cite{Boccaletti}.

To quantify the level of interdependency between two networks, we next
define the cross-clustering coefficient $C_{ij}$ for two scale-free
interdependent networks, each with $N$ nodes. Following the definition
of the clustering coefficient for a single network \cite{Watts}, we
define the cross-clustering coefficient to be
\begin{eqnarray}
C_{ij} = {\cal N}_{ij}/k_i \tilde k_j,
\label{cross}
\end{eqnarray}
where $k_i$ and $\tilde k_j$ are the number of neighbors that nodes
$i$ and $j$ have within its own network, and ${\cal N}$ the number of
links between the nodes comprising $k_i$ and $\tilde k_i$. For the
previous example of the Internet considering, e.g., the two
interdependent couples (UK-Germany), chosen because the network size for
each country is comparable, we find $\langle C_{ij} \rangle = 0.155
$. Note that for two independent BA networks, $C_{ij}$ is zero.

For the sake of simplicity we first model a NON system with only two
interdependent networks.  In model {\bf I}, each of the two
interdependent networks BA$_1$ and BA$_2$ begins with a small number
($m_0$) of nodes.  At each time step $t$, we create a new BA$_1$ node
$j$ with (i) $m_1(\le m_0$) edges that link the new node $j$ to $m_1$
already existing nodes in BA$_1$, and with (ii) $m_{12}$ edges that link
$j$ to $m_{12}$ already existing nodes in BA$_2$.  We assume that nodes
in BA$_1$ and BA$_2$ linked to $j$ are chosen based on a version of
preferential attachment---the probability $\Pi$ that a new node $j$ in
BA$_1$ is connected to node $i$ in BA$_1$ depends on the total number of
links of node $i$ with the already existing BA$_1$ and BA$_2$ nodes
(total connectivity).  Similarly, the same probability $\Pi$ controls
whether a new node $j$ in BA$_1$ is connected to node $i'$ in BA$_2$.

We define the growth of the BA$_2$ network similarly.  At each time step
$t$ we add to the BA$_2$ network a new node $j'$ with $m_2(\le m_0$)
edges that link $j'$ preferentially to $m_2$ different nodes already
present in BA$_2$ and with $m_{21}$ links that link $j'$ preferentially
to $m_{21}$ already existing nodes in BA$_1$. To reduce the number of
parameters we set $m_{21} = m_{12}$. Note that if $m_{21}=0$, while
$m_{12} \ne 0$, then due to $m_{21}=0$ each node in BA$_1$ has an equal
number of links $(m_{12})$ to nodes in BA$_2$, which is unlikely in
real-world networks.  After $t$ time steps, the four parameters of model
{\bf I}---$m_1$, $m_2$, $m_{12}$, and $m_{21}$---lead to an
interdependent network system with $t + m_0$ nodes in both BA$_1$ and
BA$_2$.  BA$_1$ has the average degree $ \langle k \rangle = 2 m_1 +
m_{12} + m_{21}$ and BA$_2$ has $ \langle k \rangle = 2 m_2 + m_{12} +
m_{21}$.  We perform numerical simulations in which $m_{21} = m_{12} $.
We then calculate the probability $P(k)$ that a node in BA$_1$ has $k$
edges either with BA$_1$ or BA$_2$ nodes.  We set $m_1 = m_2 = 3$, and
vary $m_{12}= m_{12}$.
 
\begin{figure}
\centering
\includegraphics[width=0.5\columnwidth]{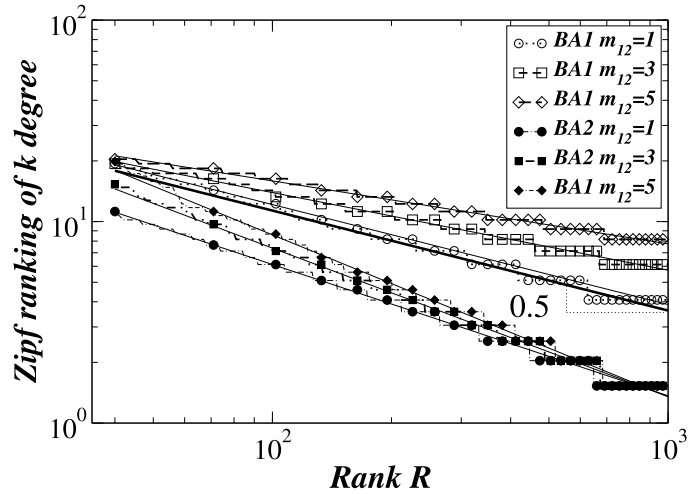}
\caption{Power law in the plot that a node has $k$ edges, for a
  model {\bf I} where $m_1 = m_2 = 3$,  $m_{21} = 1$ and $m_{12}$ 
  is varying as 1, 3,
  and 5.  Each network, i.e.  BA$_1$ and BA$_2$, has 1,000 nodes.
 The Zipf slope $\zeta$ is inverse of the cumulative distribution
  exponent $\gamma$, where $\zeta = 1 / (\gamma - 1)$.   With increasing
  $m_{12}$, the Zipf slope $\zeta$ for BA$_1$ is decreasing $(\gamma $
  increasing), whereas the Zipf slope for BA$_2$ is increasing $(\gamma
  $ decreasing).  We show the case where $m_{12} =m_{21} = 0$, $\zeta = 0.5
  ~(\gamma = 3)$, characteristic for the BA model. We show that $P(k)$ is
  characterized by a power-law exponent that is a function of the number
  of links $m_1$, $m_2$, $m_{12}$ and  $m_{21}$. 
 With increasing $m_{12}$, for BA$_1$ we have $\hat{\gamma}_1 = 
 2.776 \pm 0.006 $ ($\hat{\gamma}'_1 =3.04 \pm 0.20  $),
 $\hat{\gamma}_3 = 3.199 \pm 0.007  $ ($\hat{\gamma}'_3 = 3.32 \pm 0.23  $) 
 and 
 $\hat{\gamma}_5 = 3.56\pm 0.01 $ ($\hat{\gamma}'_5 = 3.25 \pm 0.22  $).
  With increasing $m_{12}$, for BA$_2$ we have $\hat{\gamma}_1 = 
  2.800 \pm 0.006 $ ($\hat{\gamma}'_1 = 3.09 \pm 0.20  $),
 $\hat{\gamma}_3 = 2.541 \pm 0.005 $ ($\hat{\gamma}'_3 = 2.66 \pm 0.17  $) 
 and 
 $\hat{\gamma}_5 = 2.343 \pm 0.005  $ ($\hat{\gamma}'_5 = 2.52 \pm 0.15  $)
   }
\label{fig2}
\end{figure}

Figure~\ref{fig2} shows that, when $m_{21} = 1$, the Zipf plot of $k$
exhibits a power law for varying values of $m_{12}$.  With increasing
$m_{12}$, $\zeta$ of the Zipf plot decreases ($\gamma$ of $P(k)$
increases), and the $\gamma$ exponent for BA$_2$ decreases. When $m_{12}
= 0$, BA$_1$ and BA$_2$ become decoupled and yield $(\gamma = 3)$, which
is characteristic of the BA model. Thus the power-law exponent $\gamma$
of $P(k)$ is a function of the number of links $m_1$, $m_2$, and
$m_{12}(m_{21})$ and, due to interdependencies, $\gamma$ can change
substantially for different networks.

Next, for model {\bf I} we present analytic solutions for the power-law
exponent $\gamma$ of $P(k)$ as a function of the number of links both
between and within networks.  We apply the continuum approach introduced
in Refs.~\cite{Albert00,AlbertRMP}, which calculates the time dependence
of the degree of a given node $i$, e.g., for BA$_1$.  $k^T_{1,i}$ is the
total number of edges between $i$ in BA$_1$ and other nodes in
BA$_1$---$k_{1,i}$---and between $i$ and nodes in BA$_2$---$k_{21,i}$,
\begin{eqnarray}
k^T_{1,i} = k_{1,i} + k_{21,i}.
\label{BP10} 
\end{eqnarray}
The probability that a new node $j$ created in BA$_1$ will link to an
already existing node $i$ in BA$_1$ depends on the probability of this
process, $\Pi(k^T_{1,i})$.  Approximating $k^T_{1,i}$ with a continuous
real variable \cite{AlbertRMP}, the rate at which $k^T_{1,i}$ changes we
expect to be proportional to $\Pi(k^T_{1,i})$ where
\begin{eqnarray}
\frac{\partial k^T_{1,i}}{\partial t} = (m_1+ m_{21}) \Pi(k^T_{1,i}) = 
\frac{(m_1+ m_{21}) k^T_{1,i}}{2 m_1 t + m_{12} t + m_{21} t }.
\label{BP1} 
\end{eqnarray}
>From the denominator in the last expression we note that each endpoint
of an $m_1$ edge is a node in BA$_1$ because $m_1$ edges are established
between nodes in BA$_1$. This is in contrast to $m_{21}(m_{12})$ edges
where one end is linked to a node in BA$_1$ and the other to a node in
BA$_2$.  The initial condition is that every new node $i$ must have a
degree $k^T_{1,i}(t_i) = m_1 + m_{12}$, since it connects to $m_1$ nodes
in BA$_1$ and $m_{12}$ in BA$_2$.  From Eq.~(\ref{BP1}), we obtain
\begin{eqnarray}
 k^T_{1,i}(t) &=& (m_1 +  m_{12}) ({t}/{t_i})^{\beta_1}, \mathrm{~where}\nonumber\\
 \beta_1 &=& \frac{m_1+ m_{21}}{2 m_1  +  m_{12} +  m_{21}}.
\label{kt}
\end{eqnarray}
Note that in the limiting case $m_{12} = m_{21} = 0$ the networks
decouple with $\beta = 1/2$, as in the BA model
\cite{Barabasi99,AlbertRMP}. Other choices for $\beta$ in single
networks are proposed in different models
\cite{Krapivsky00,Dorogovtsev00,AlbertPRL00}.

The probability that a node $i$ has a degree $k^T_{1,i}(t_i)$ smaller
than $k^T$ is \cite{Albert00,AlbertRMP}
\begin{equation}
  P[k^T_{1,i}(t) < k^T] = P\left[ t_i > \frac{(m_1 + m_{12})^{1/\beta_1} t}
  {(k^T)^{1/\beta_1}}\right].  
\label{pdf1} 
\end{equation}
Assuming that new nodes are entered homogeneously in time, the
distribution of $t_i$ values is $P(t_i) = 1 / (m_0 + t)$. Entering this
expression into Eq.~(\ref{pdf1}) we obtain
%
$P( t_i > \frac {\big(m_1 + m_{12}\big)^{1/\beta_1} t}
{(k^T)^{1/\beta_1}}) = 1 -   
\frac{(m_1 + m_{12})^{1/\beta_1} t}{(k^T)^{1/\beta_1} (t + m_0)}, $
%
and the degree distribution $P(k^T)$ of BA$_1$ 
\begin{equation}
P(k^T) =  \frac {\partial P(k_{1,i} < k^T)}{\partial k^T} = 
\frac{ (m_1 + m_{12})^{1/\beta_1} t}{(k^T)^{1/\beta_1 + 1} (t + m_0) \beta_1}, 
\label{BA3} 
\end{equation} 
where, asymptotically, for $t \rightarrow \infty$ (networks with an
infinite number of nodes), the above equation yields
\begin{equation}
   P(k^T)  \propto (k^T)^{-\gamma_1}, {\rm ~where}~~~~\gamma_1 =
   \frac{1}{\beta_1} + 1, 
\label{gamma1} 
\end{equation}
with $\beta_1$ defined as in Eq.~(\ref{kt}).  Similar to
Eq.~(\ref{BP1}), $k^T_{2,i}$ is the total number of links for a node $i$
in BA$_2$, which is the total number of edges between BA$_2$ node $i$
and other nodes in both BA$_1$ and BA$_2$, and satisfies the dynamic
equation
%
$ \frac{\partial k^T_{2,i}}{\partial t} =  (m_2+ m_{12}) 
\Pi(k^T_{2,i}) = \frac{(m_2+ m_{12}) k_{2,i}}{2 m_2 t + m_{12} t + m_{21} t }.$
%
Following Eqs.~(\ref{pdf1})-(\ref{BA3}), the degree distribution
$P(k)$ in the BA$_2$ network, $\gamma_2$, and $\beta_2$ is similar to
that in Eqs.~(\ref{kt}) and (\ref{gamma1}) in which 1 is replaced by 2
and vice versa.

Unlike the pure BA model, in which $\beta = 1/2$
\cite{Barabasi99,AlbertRMP}, in the coupled BA model we find that the
power-law exponent of the degree distribution depends on the number of
edges within each network, $m_1 (m_2)$, and on the number of edges
between the interdependent networks $m_{12} (m_{21})$.  Also, in
agreement with Fig.~1, when $m_{21} = 0$, for each $m_{12}$, $\beta_1
\le 0.5$ implies $\gamma_1 \ge 3$ for $P(k)$, whereas $P(k)$ for BA$_2$
has $\gamma_1 \le 3$.

In addition to the degree distribution for the total number of links
$k_i^T$ of Eq.~(\ref{kt}), we next provide an analytic result for the
degree distribution for the number of links between nodes within a
BA$_1$ network. Following Eqs.~(\ref{BP10}) and (\ref{BP1}), we obtain
%
 $\frac{\partial k_{1,i}}{\partial t} =  
  \frac{m_1 ~k^T_{1,i}}{2 m_1 t + m_{12} t + m_{21} t }$.
%
Entering Eq.~(\ref{kt}) into the previous equation, we obtain
%
$k_{1,i}(t) = \frac{m_1 (m_1 + m_{12})}{m_1 + m_{21} } 
\big(\frac{t}{t_i}\big)^{\beta_1} + \frac{m_1 (m_{21} -m_{12})}{m_1 + m_{21}}$.
%
Following Eqs.~(\ref{pdf1})--(\ref{BA3}), the degree distribution
$P(k)$ for the total number of links between nodes within network BA$_1$
scales as $P(k) \propto k^{-\gamma_1}$ for $t \rightarrow
\infty$. Similarly, we calculate the degree distribution $P(k)$ for the
total number of links between different networks and again obtain $P(k)
\propto k^{-\gamma_1}$ where $k_{21,i}(t) = \frac{m_{21} (m_1 +
  m_{12})}{m_1 + m_{21} } \left(\frac{t}{t_i}\right)^{\beta_1} +
\frac{m_{21} (m_{21} -m_{12})}{m_1 + m_{21}}$.  Thus the scaling
exponent for $P(k)$ is the same for links connecting nodes of different
networks, $k_{21,i}(t)$, links within a given network, $k_{1,i}(t)$, and
for the total number of links, $k_{1,i}^T(t)$.  In practice, by testing
this regularity we can determine whether a given pair of interdependent
networks follows model {\bf I}.

Model {\bf I} has two interesting limits, (i) when $m_{12} = m_{21} =
m^I$, $\beta_1 = \beta_2 = 1/2$, as in the pure BA model, and (ii) when
$m_{12} \rightarrow \infty$ nodes of BA$_1$ establish many more
connections with BA$_2$ than with other nodes in BA$_1$.  This implies
that $\beta_1 \rightarrow 0$, as in Eq.~(\ref{kt}), and $\beta_2
\rightarrow 1$, which yields exponents $\gamma_1 \rightarrow \infty$
(the Gaussian limit), as in Eq.~(\ref{gamma1}), and $\gamma_2
\rightarrow 2$ (the Zipf law).

\begin{figure}
\centering
\includegraphics[width=0.5\columnwidth]{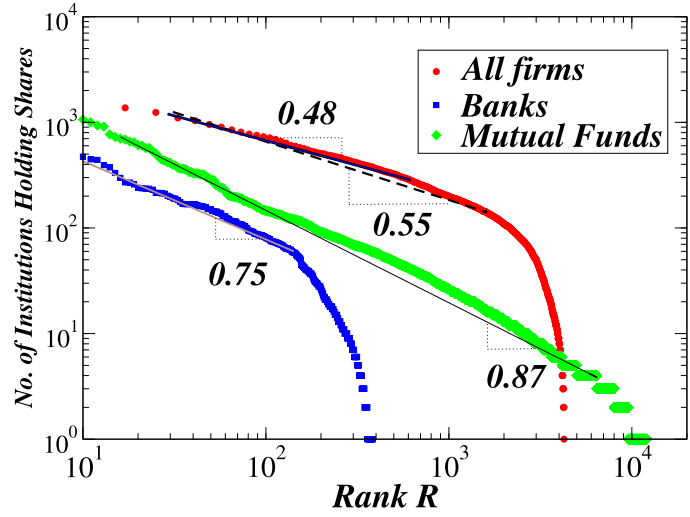}
\caption{Power laws in interdependent financial networks.  
Power law in the Zipf plot with exponent $\zeta = (1 - \gamma)$ for total number of links
 versus  rank $R$ for  $15, 636$ mutual funds, $4,544$
 US firms and separately for 384 US banks. For each firm $i$ we 
calculate  links from  mutual funds and other
 firms to firm  $i$.  
  For each mutual funds $i$ we 
calculate  links from  mutual fund $i$ to 
  other mutual funds. 
 We obtain the following exponents:
 for firms $\hat{\gamma} = 2.725 \pm 0.008 $ ($\hat{\gamma}' =
  3.42 \pm 0.17   $);
 for banks $\hat{\gamma} = 2.17 \pm 0.02 $ 
 ($\hat{\gamma}' = 2.39 \pm 0.31  $); for 
 mutual funds $\hat{\gamma} = 2.231 \pm 0.002 $ ($\hat{\gamma}' = 
 2.31 \pm 0.09 $).}
\label{fig3}
\end{figure}

We further exemplify the utility of model {\bf I} using two networks
from {\sc Yahoo Finance\/} for 2011.  Figure~\ref{fig3} shows 4,544 US
firms listed on the NYSE and Nasdaq representing network BA$_1$, and
15,636 mutual funds representing network BA$_2$. Note that firms
comprising BA$_1$ and mutual funds comprising BA$_2$ present only a
partial picture of the complete financial network. Clearly, one may
extend this analysis by including additional networks such as hedge
funds and pension funds. For each firm $i$ of BA$_1$ we show the total
number of holders, i.e., the total number of institutions holding shares
(including links from institutional owners such as pension funds, banks,
mutual funds, and hedge funds, but also other firms linked to $i$),
$k_{1,i}^T$. Thus because mutual funds comprising BA$_2$ hold shares in
BA$_1$, interdependency between the two networks is established.
Figure~\ref{fig3} shows the exponents of
Eqs.~(\ref{gabaix})-(\ref{alpha}) for US firms: $\hat{\gamma} = 2.73 \pm
0.01 $ ($\hat{\gamma}' = 3.42 \pm 0.17 $).  For each mutual fund $i$ of
BA$_2$ we show the total number of holdings, which includes firms of
BA$_1$ and also pension funds and other institutions not included in our
study.  We show the exponents of Eqs.~(\ref{gabaix})--(\ref{alpha}) for
mutual funds: $\hat{\gamma} = 2.23 \pm 0.002 $ ($\hat{\gamma}' = 2.31
\pm 0.09$). Figure~\ref{fig3} shows the plot $k^T_{1,i}$ vs. rank
between rank 20 and 2000.  Figure~\ref{fig3} also shows $k^T_{1,i}$
vs. rank for US banks, which represent only a small fraction of the
total number of US firms, where $\hat{\gamma} = 2.17 \pm 0.02 $
($\hat{\gamma}' = 2.39 \pm 0.31$).  We note that we can replicate these
diverse values for $\gamma_1$ and $\gamma_2$ using model {\bf I}.

\begin{figure}
\centering
\includegraphics[width=0.5\columnwidth]{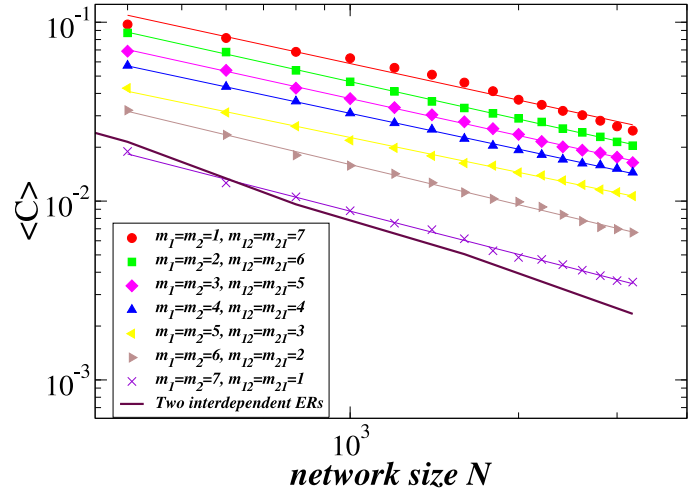}
\caption{Power law in the cross-clustering coefficient versus size of
  the two interdependent   Baraba\'si-Albert (BA) model with $\langle k
  \rangle = 16$, compared with the cross-clustering coefficient of a
  random graph, $\approx N^{-1} $.  With increasing $m_{12} = m_{21}$
  the intercept of power law increases. 	
 }
\label{fig4}
\end{figure}

Next we study the scaling of the cross-clustering coefficient $C_{ij}$
of Eq.~(\ref{cross}) for two scale-free interdependent networks, each
with $N$ nodes, as a function of system size.  We study the average of
$C_{ij}$ versus $N$, $\langle C \rangle $ versus $N$.  To give context
to $\langle C \rangle $: in a friendship network $\langle C \rangle $
reflects to what extent an $i$-friend from city A and another
$i$-friend from city B know each other. Figure~\ref{fig4} fixes $\langle k \rangle =
16$, and varies $m_1$, $m_2$, and $m_{12} = m_{21} $ in order to
numerically determine that $\langle C \rangle $ vs. $N$
follows a power law with an average slope $0.71 \pm 0.02$, a value close
to $0.75$, which is also obtained numerically for the global cluster
coefficient for a single BA network \cite{AlbertRMP}. As $m_{12} =
m_{21}$ increases, the intercept of $\langle C \rangle$ vs. $N$ also
increases.  Note that for two independent BA networks $\langle C
\rangle$ is zero for all $N$.  We also study two interdependent
Erdos-Renyi (ER) networks, A and B, each of size $N$, where the
probability of all links, both between and within networks, is
$p$. First we find numerically that $p=0.5\cdot\langle k\rangle/(N-1)$
is needed in order to reproduce a given $\langle k\rangle$ (note that
$p=\langle k\rangle/(N-1)$ corresponds to a single ER network). We next
find that the cross-clustering coefficient $\langle C \rangle $ vs. $N$
also follows a power law with slope $-1$, the same slope as found for
the clustering coefficient vs. $N$ in a single ER model
\cite{AlbertRMP}. Figure~\ref{fig4} shows that the cross-clustering
coefficient $\langle C \rangle$ for two interdependent BA models is
stronger than $\langle C \rangle$ for two interdependent ER models.
 
In order to define a new scale-free interdependent network model {\bf
  II} in which we separately define the dynamics for growing links
within a network and the dynamics for growing links between networks.
In model {\bf II} we create a new BA$_1$ node $j$ with $m_1$ edges that
link $j$ to $m_1$ existing nodes in BA$_1$, and with $m_{12}$ edges that
link $j$ to $m_{12}$ existing nodes in the BA$_2$ network at each $t$.
Similarly, we link a new node $j'$ created in BA$_2$ with $m_2$ edges to
$m_2$ existing nodes in BA$_2$. We link new node $j'$ to $m_{21}$
existing nodes in BA$_1$.  Links within networks, $k_{1,i}$ and
$k_{2,i}$, are treated according to the ordinary scale-free BA model,
i.e., using the continuum approach \cite{AlbertRMP}
$\frac{\partial k_{1,i}}{\partial t}  =  m_1 \Pi(k_{1,i})  =  \frac{m_1  k_{1,i}}{2 m_1 t }$ and 
$\frac{\partial k_{2,i}}{\partial t}  =  m_2 \Pi(k_{2,i})  =  \frac{m_2  k_{2,i}}{2 m_2 t }$.
%
Thus links within a network only attract new links created within the
same network. We similarly define that only links between networks can
attract new links established between networks. The number of links
of BA$_1$ node $i$ with nodes in BA$_2$, $k_{21,i}$, and the number of
links of BA$_2$ node $i$ with nodes in BA$_1$,$k_{12,i}$, satisfy
%
 $\frac{\partial k_{21,i}}{\partial t} = m_{21} ~ \Pi(k_{21,i}) 
  =  \frac{m_{21} ~ k_{21,i}}{ m_{21} t }$ and  
%
%
$ \frac{\partial k_{12,i}}{\partial t} = m_{12} ~ \Pi(k_{12,i}) 
  =  \frac{m_{12} ~ k_{12,i}}{ m_{12} t }.$
%
Note that in edges $m_{21}(m_{12})$, one end is linked to a node in
BA$_1$ and the other to a node in BA$_2$.  Following
Eqs.~(\ref{BP1})--(\ref{gamma1}), we find that the degree distribution
$P(k)$ of the number of links between BA$_1$ and BA$_2$ becomes $P(k)
\propto k^{-\gamma_3}$ where $\gamma_3 = \frac{1}{\beta_3} + 1$ and
$\beta_3 = 1$.  This demonstrates that the power-law exponent $\gamma_3$
of $P(k)$ does not depend on parameters $m_1$, $m_2$, $m_{12}$, and
$m_{21}$.  In addition, $P(k)$ follows a Zipf law. In practice, we can
determine whether a pair of interdependent networks follows model {\bf
  II} by testing this regularity.


Models {\bf I} and {\bf II}, which we have used to study network pairs,
can be generalized to $N$ interdependent networks.  For each pair
$(I,J)$ where $I$ and $J$ run from 1 to $N$, at each time step $t$ we
add a new node $j$ to BA$_I$ with $m_I$($\le m_0$) edges to $m_I$
already existing nodes in BA$_I$ and $m_{IJ}$ edges to $m_{IJ}$ nodes
already existing in BA$_J$.  Applying Eqs.~(\ref{BP10})--(\ref{BP1}),
defined for a pair of networks, to the $N$ networks case (the NON model),
for $k_{I,i}^T$---the total number of edges between a node $i$ and other
nodes in BA$_I$, and between $i$ and other nodes in BA$_J$---we obtain
%
$ \frac{\partial k^T_{I,i}}{\partial t} 
  = \frac{(m_I+ \Sigma_{J=1}^N m_{JI}) k^T_{I,i}}{2
   m_I t + \Sigma_{J=1}^N m_{IJ} t +\Sigma_{J=1}^N m_{JI} t }.$
%
Following Eqs.~(\ref{BP1})-(\ref{gamma1}), we find that the degree
distribution $P(k)$ of the number of links between BA$_I$ and BA$_J$
becomes $P(k) \propto k^{-\gamma_5}$ where $\gamma_5 = \frac{1}{\beta_5}
+ 1$, and $\beta_5 = (m_I+\Sigma_{J=1}^N m_{JI})/(2 m_I + \Sigma_{J=1}^N
m_{IJ} + \Sigma_{J=1}^N m_{JI}).$

Understanding the dynamics of interdependent networks---how different
networks simultaneously evolve in time---is a necessary precondition to
predicting the behavior of networks over time, and to discovering how
quickly failures initiated in one network spread to other networks
\cite{Schweitzer09,Vespignani}.


\acknowledgments
This study is  supported by The National Science
Foundation, DTRA.

%
%
%

\bigskip 








\newpage  

\end{document}